# 'Propeller' driven spectral state transition in LMXB 4U 1608-52

Xie Chen[1]

Shuang Nan Zhang[1,2,3,4]

Guo Qiang Ding[1]

[1]Physics Department and Center for Astrophysics, Tsinghua University, Beijing, 100084, China (zhangsn@tsinghua.edu.cn, dinggq@mail.tsinghua.edu.cn, carolchenxie@gmail.com)

[2]Key Laboratory of Particle Astrophysics, Institute of High Energy Physics, Chinese Academy of Sciences, Beijing, China

[3]Physics Department, University of Alabama in Huntsville, Huntsville, AL35899, USA (zhangsn@uah.edu)

[4]Space Science Laboratory, NASA Marshall Space Flight Center, SD50, Huntsville, AL35812, USA

Abstract

Spectral state transitions in neutron star LMXB systems have been widely observed yet not well understood. Here we report an abrupt spectral change in 4U 1608-52, a typical atoll source, during its decay phase of the 2004 outburst. The source is found to undergo sudden changes in its spectral hardness and other properties. The transition occurred when its luminosity is between $(3.3 - 5.3) \times 10^{36}$ $ergs$ $s^{-1}$, assuming a distance of 3.6 kpc. Interpreting this event in terms of the 'propeller effect', we infer the neutron star surface magnetic field as $(1.4\text{-}1.8) \times 10^8$ Gauss. We also briefly discuss similarities and differences between the spectral states of neutron star and black hole binary systems.

*Subject headings*: accretion, accretion disks—binaries: general—stars: individual (4U1608-52)—stars: neutron

1. Introduction

One class of neutron star Low-Mass-X-ray-binaries, the atoll sources, are found to follow an atoll-shaped track on the Color-Color diagram (Hasinger & van der Klis, 1989). In the 'banana' part, a curved branch to the bottom and the right sides of the diagram, the sources are in a high soft state, while in the 'island' part, the isolated patches to the top of the diagram, their X-ray spectra are low and hard. Most of previous studies tracking the spectral evolution of atoll sources between these two states suggested that mass accreting rate $\dot{M}$ on to the neutron star from its companion is the driving force behind (see for example Bloser, et al., 2000; Gierlinski M., Done C., 2002 ). They argued that $\dot{M}$ determines the inner truncation radius of the accretion disk and thus the temperature of the soft component and the hardness of the spectra. However it is still not well understood exactly what drives the transition, as state transition has been found taking place at very different luminosities in similar sources (Bloser et al, 2000). Detailed analysis of continuous spectral evolution around the abrupt jumps from high/soft state to low/hard state in some LMXBs has, within the above framework, attributed the phenomena further to the centrifugal barrier, the so called 'propeller' effect (Zhang et al., 1998).

4U 1608-52 is a typical atoll source (Straaten, et al, 2003) from which highly coherent 619 Hz oscillations in thermonuclear X –ray bursts have been detected (Hartman, et al, 2003). This leads to a determination of neutron star spin period to be 1.61 ms. Its distance is estimated to be 3.6 kpc from observations of flux-saturated type-I X-ray bursts (Nakamura, et al, 1989). Following its atoll-shaped path, the spectrum evolves from high soft state to low hard state and back again. Fitting it with a black body plus comptonization model has generally been successful (Gierlinski M., Done C., 2002) but detailed spectral analysis is still needed to answer questions about the underlying mechanism of its evolution process. Zhang et al. (1996) first detected its hard X-ray outburst between 20-100 keV with the CGRO/BATSE instrument (Zhang et al. 1994, Harmon et al. 2002); similar hard X-ray outburst has also been detected from a neutron star X-ray binary Aquila X-1 with CGRO/BATSE (Harmon et al. 1996).

Continuous RXTE (Bradt, Rothschild & Swank 1993) broadband X-ray observations of 4U 1608-52 have made it possible for us to study its spectral evolution more closely. In this paper, we report detection of a sudden spectral change observed in 4U 1608-52. By accounting for it with the propeller effect, we can estimate the neutron star surface magnetic field. Energy spectra of BH binaries are also seen to undergo high/soft to low/hard transitions. Comparison of 4U 1608-52 with BH binaries is also carried out here in search of traces of different effects of their central compact objects.

2. Observations and analysis

We analyzed 4U 1608-52 data in RXTE observations through 2004 March 19th to 2004 April 26th, the decay phase of the 2004 March outburst. All data between March 19th and April 7th are included. (A type-I X-ray burst is present on March 26th and we excluded data 700 seconds before and after it). However, for a long period after April 7th, observations were too short to yield any reliable spectral information and therefore excluded from our analysis. We sampled three more observations from April 19th to April 26th to investigate spectral properties of the source when it had decayed further.

2.1 Light curve and Color-Color Diagram

We used Standard2f mode PCA data to plot light curve and color-color diagram for all observations. Background files were generated using latest standard PCA background models and subtracted to produce net light curves. Figure 1 shows the overall count rate of PCA in the 3.0-25.0 keV energy band for each observation. Within the overall decreasing trend, we can divide the observations into three intensity subgroups: the first four with count rate above 600 counts/s, the next 12 below 600 counts/s but above 150 counts/s, and the last three below 150 counts/s.

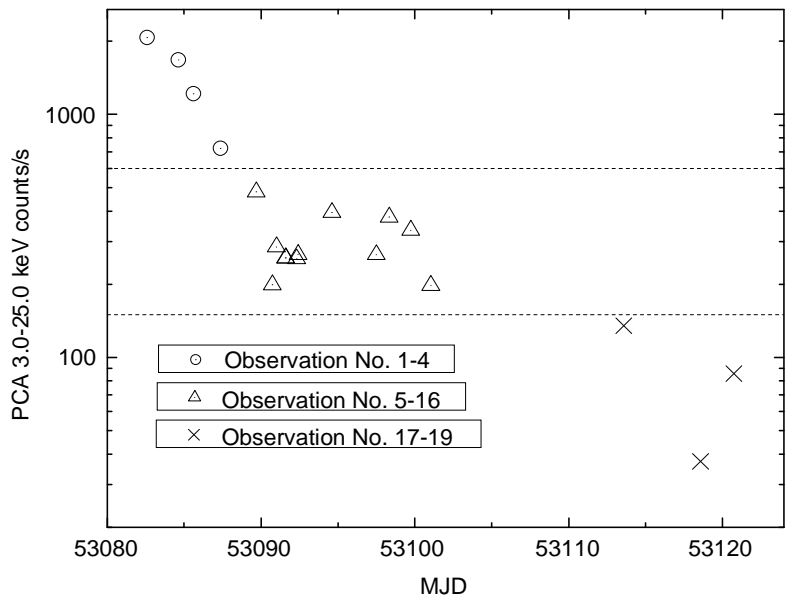

Fig 1.—light curve for the decay phase of the 2004 outburst. Total count rate for PCA data of each observation in the 3.0-25.0 keV energy band is drawn vs. the observation starting time. The upper dashed line corresponds to count rate=600 counts/s, and the lower corresponds to count rate=150 counts/s.

When plotting the Color-Color diagram, soft color is defined as the count rate ratio between 4.0 to 6.5 keV and 2.9 to 4.0 keV, and hard color between 9.8 to 16.0 keV and 6.5 to 9.8 keV. Every point on the CCD (Fig. 2) represents a 16 second time interval. We use the corresponding symbols to represent data in the three groups defined above. An abrupt change in hard color between the 4th and the 5th observation is obvious. Both before and after the jump the position of the source on the CCD remains rather stable. Between observation 16 and 17, there is a time gap of 12 days. The flux dropped by more than 50% but the source hardly moved on the CCD. It only exhibited greater dispersion on the graph which should be attributed to the relatively higher counting error when the flux is low.

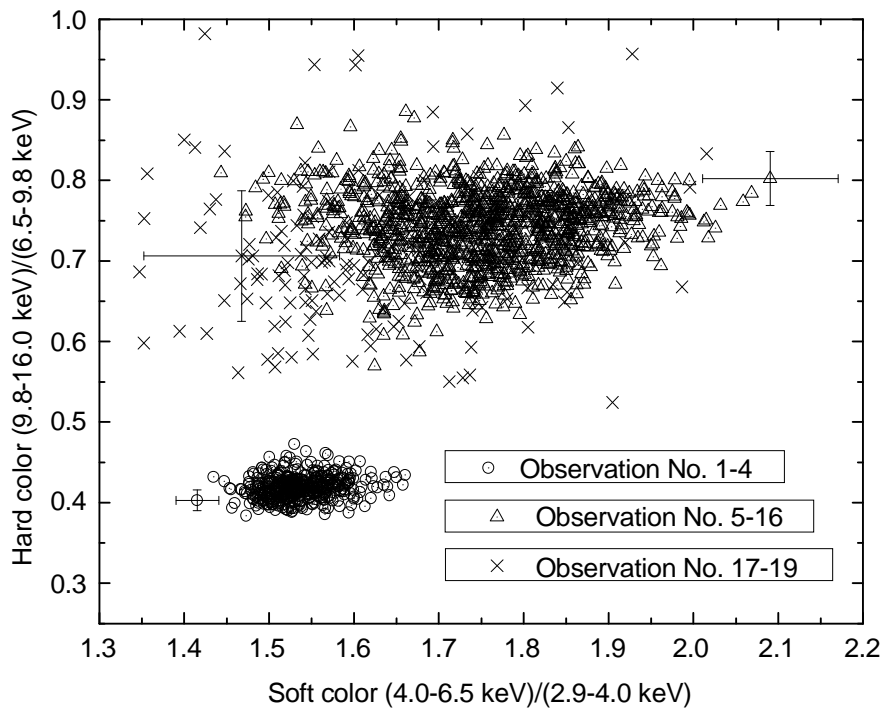

Fig 2.—Color-color diagram for the 19 observations. Hard and soft colors are defined as count rate ratios in the (9.8-16.0) / (6.5-9.8) keV and (4.0-6.5) / (2.9-4.0) keV band respectively. Each point in the diagram spans 16 second. Data in the three groups are represented with different symbols: circle for the first group, triangle for the second and cross for the third. One set of error bar is plotted for each group respectively. It is clear that the error in the colors increases when the flux drops.

2.2 Spectral analysis

Spectrum for each observation is constructed by combining PCA and HEXTE data together. We use 3.0-25.0 keV Standard2f mode PCA data and Cluster 0 archive data of HEXTE between 15.0 and 150.0 keV. A 0.3% systematic error is added to PCA data. Though this is smaller than the usual practice (0.5%-1%), the fitting result is already satisfactorily good. Count rate above 30 keV is low so we regrouped energy bins of HEXTE into larger ones in order to have better statistics. While fitting the two data sets jointly, we leave the relative normalization between them as a free parameter to account for relative instrumental calibration uncertainties.

We have fitted all the spectra successfully with a two-component model in XSPEC: a power law with an exponential cutoff (an analytical approximation of unsaturated Comptonization) plus a blackbody. An iron line feature around 6.7 keV is also included, which greatly improved the fitting outcome. The fitting result of the 19 observations are listed in Table 1. Flux of the blackbody and the cutoff powerlaw component is calculated over the entire 3.0—150.0 keV energy range. While the blackbody temperature $kT$ stayed quite stable (around 0.7 keV) in all observations, the power law index $a$ and cutoff energy $Ecut$ change dramatically. It is clear in figure 3 that an abrupt spectral state transition occurred between the 4$^{th}$ and 5$^{th}$ observation as $Ecut$ changed from ∼3keV to above 30 keV.

Table 1
Spectral Fitting Results of all observations with the BB+CPL model

| Observation No. | Power law | | | Black Body | | Iron line | | | $\chi^2$/47 |
| | Index | Ecut | Flux | kT | FLux | LineE | Sigma | Norm | |
| (1) | (2) | (3) | (4) | (5) | (6) | (7) | (8) | (9) | (10) |
| 1 | $0.17^{+0.09}_{-0.03}$ | $3.42^{+0.04}_{-0.04}$ | 4.5844E-9 | $0.72^{+0.01}_{-0.01}$ | 1.2377E-9 | $6.87^{+0.20}_{-0.20}$ | $1.14^{+0.17}_{-0.15}$ | $1.17E-02^{+0.34E-02}_{-0.20E-02}$ | 1.487 |
| 2 | $0.22^{+0.08}_{-0.03}$ | $3.25^{+0.03}_{-0.03}$ | 3.6123E-9 | $0.66^{+0.01}_{-0.01}$ | 9.3176E-10 | $6.83^{+0.10}_{-0.13}$ | $1.09^{+0.18}_{-0.11}$ | $1.14E-02^{+0.20E-02}_{-0.31E-02}$ | 1.961 |

| (1) | (2) | (3) | (4) | (5) | (6) | (7) | (8) | (9) | (10) |
|---|---|---|---|---|---|---|---|---|---|
| 3 | $0.34^{+0.09}_{-0.04}$ | $3.55^{+0.05}_{-0.24}$ | 3.7287E-9 | $0.68^{+0.02}_{-0.01}$ | 7.0549E-10 | $6.90^{+0.15}_{-0.40}$ | $1.27^{+0.18}_{-0.35}$ | $1.02E-02^{+0.36E-02}_{-0.22E-02}$ | 1.103 |
| 4 | $1.2^{+0.4}_{-0.4}$ | $5.97^{+0.20}_{-0.95}$ | 2.2639E-9 | $0.67^{+0.02}_{-0.03}$ | 3.6905E-10 | $6.69^{+0.12}_{-0.24}$ | $1.66^{+0.15}_{-0.11}$ | $1.81E-02^{+0.48E-02}_{-0.85E-02}$ | 1.880 |
| 5 | $1.75^{+0.01}_{-0.01}$ | $>174$ | 3.3269E-9 | $0.91^{+0.03}_{-0.04}$ | 1.1953E-10 | $7.06^{+0.09}_{-0.09}$ | $0.90^{+0.12}_{-0.11}$ | $2.24E-03^{+0.31E-03}_{-0.20E-03}$ | 1.448 |
| 6 | $1.45^{+0.03}_{-0.03}$ | $55.0^{+12.5}_{-9.1}$ | 2.4009E-9 | $0.80^{+0.02}_{-0.02}$ | 1.3810E-10 | $6.95^{+0.18}_{-0.15}$ | $0.77^{+0.20}_{-0.14}$ | $1.62E-03^{+0.66E-03}_{-0.33E-03}$ | 1.434 |
| 7 | $1.39^{+0.03}_{-0.04}$ | $55.6^{+11.5}_{-9.3}$ | 2.5228E-9 | $0.85^{+0.02}_{-0.02}$ | 1.4804E-10 | $6.93^{+0.16}_{-0.16}$ | $0.76^{+0.19}_{-0.17}$ | $1.35E-03^{+0.60E-03}_{-0.30E-03}$ | 0.925 |
| 8 | $1.40^{+0.02}_{-0.05}$ | $48.0^{+12.3}_{-9.6}$ | 2.1233E-9 | $0.83^{+0.03}_{-0.03}$ | 1.2602E-10 | $6.92^{+0.20}_{-0.19}$ | $0.48^{+0.28}_{-0.27}$ | $9.25E-04^{+4.92E-04}_{-2.75E-04}$ | 0.795 |
| 9 | $1.52^{+0.07}_{-0.04}$ | $87.2^{+33.9}_{-20.8}$ | 2.4335E-9 | $0.83^{+0.02}_{-0.03}$ | 1.1407E-10 | $7.04^{+0.20}_{-0.19}$ | $0.72^{+0.23}_{-0.20}$ | $1.05E-03^{+0.52E-03}_{-0.26E-03}$ | 1.155 |
| 10 | $1.41^{+0.03}_{-0.08}$ | $57.4^{+13.6}_{-8.5}$ | 2.2619E-9 | $0.77^{+0.02}_{-0.02}$ | 1.2246E-10 | $6.89^{+0.17}_{-0.19}$ | $0.96^{+0.22}_{-0.19}$ | $1.53E-03^{+0.74E-03}_{-0.34E-03}$ | 1.149 |
| 11 | $1.63^{+0.03}_{-0.04}$ | $92.7^{+46.0}_{-24.0}$ | 2.2090E-9 | $0.78^{+0.02}_{-0.03}$ | 1.1626E-10 | $6.90^{+0.19}_{-0.19}$ | $0.75^{+0.24}_{-0.21}$ | $1.40E-03^{+0.63E-03}_{-0.30E-03}$ | 1.046 |
| 12 | $1.48^{+0.02}_{-0.03}$ | $70.0^{+9.3}_{-9.5}$ | 3.6440E-9 | $0.91^{+0.01}_{-0.02}$ | 1.5868E-10 | $6.88^{+0.11}_{-0.11}$ | $0.57^{+0.15}_{-0.14}$ | $1.39E-03^{+0.45E-03}_{-0.26E-03}$ | 1.125 |
| 13 | $1.57^{+0.03}_{-0.04}$ | $82.2^{+29.7}_{-21.7}$ | 3.4742E-9 | $0.87^{+0.03}_{-0.04}$ | 1.3619E-10 | $6.92^{+0.20}_{-0.17}$ | $0.36^{+0.15}_{-0.26}$ | $1.23E-03^{+0.53E-03}_{-0.33E-03}$ | 1.272 |
| 14 | $1.69^{+0.03}_{-0.03}$ | $98.4^{+50.2}_{-29.4}$ | 3.0396E-9 | $0.86^{+0.03}_{-0.04}$ | 1.4294E-10 | $6.82^{+0.08}_{-0.16}$ | $0.64^{+0.24}_{-0.18}$ | $2.08E-03^{+0.78E-03}_{-0.44E-03}$ | 1.078 |
| 15 | $1.51^{+0.02}_{-0.03}$ | $68.7^{+10.0}_{-9.8}$ | 2.9151E-9 | $0.91^{+0.01}_{-0.02}$ | 1.4732E-10 | $6.95^{+0.10}_{-0.10}$ | $0.65^{+0.13}_{-0.12}$ | $1.46E-03^{+0.42E-03}_{-0.24E-03}$ | 1.586 |
| 16 | $1.64^{+0.02}_{-0.02}$ | $130.1^{+60.0}_{-37.7}$ | 2.6903E-9 | $0.85^{+0.03}_{-0.03}$ | 1.0487E-10 | $6.95^{+0.11}_{-0.11}$ | $0.49^{+0.13}_{-0.13}$ | $1.30E-03^{+0.36E-03}_{-0.23E-03}$ | 1.025 |
| 17 | $1.73^{+0.02}_{-0.02}$ | $>89.12$ | 1.2388E-9 | $0.75^{+0.07}_{-0.06}$ | 4.5028E-11 | $6.92^{+0.24}_{-0.25}$ | $<0.6806$ | $5.27E-04^{+2.52E-04}_{-3.57E-04}$ | 0.964 |
| 18 | $1.46^{+0.12}_{-0.03}$ | $>80.21$ | 6.8196E-10 | $0.74^{+0.06}_{-0.11}$ | 2.4438E-11 | $7.15^{+0.51}_{-0.50}$ | $0.88^{+0.77}_{-0.37}$ | $4.37E-04^{+6.77E-04}_{-2.20E-4}$ | 0.873 |
| 19 | $1.90^{+0.04}_{-0.02}$ | $>92.56$ | 9.7124E-10 | $0.79^{+0.03}_{-0.06}$ | 3.4090E-11 | $6.97^{+0.17}_{-0.17}$ | $0.42^{+0.25}_{-0.25}$ | $5.87E-04^{+1.78E-04}_{-3.21E-04}$ | 0.826 |

Table 1. Fitting results of the Bbodyrad+Cutoffpl+Gaussian model for all 19 spectra. PCA data between 3.0-25.0 keV and HEXTE data between 15.0-150.0 keV are used. *NH* is fixed at 1.5 E 22 cm$^{-2}$. Col. (1): Observation number. Col. (2): power law index. Col. (3): units of keV. Col. (4): units of ergs cm$^{-2}$ s$^{-1}$. Col. (5): units of keV. Col. (6): units of ergs cm$^{-2}$ s$^{-1}$. Col. (7): units of keV. Col. (8): units of keV. Col. (9): units of photons keV$^{-1}$ cm$^{-2}$ s$^{-1}$. Col. (10): reduced chi-squared for 47 degrees of freedom.

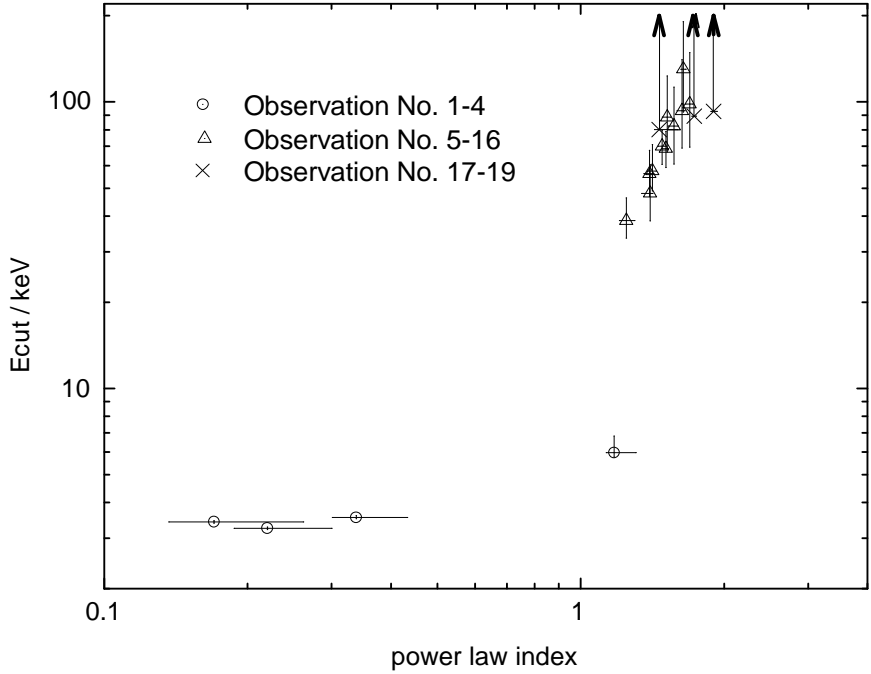

Fig 3.—Spectral fitting parameters, power law cutoff energy vs. power law index for all 19 observations. The symbols correspond to those in previous figures (crosses representing the last three observations are plotted at the lower limit of their cutoff energy). Error bars at 90% confidence level are drawn. Vertical arrows mean that the upper limit of Cutoff energy exceeds 200 keV and cannot be determined from our data.

3. Discussion and Conclusion

3.1 Interpretation with the propeller effect

From the above analysis, a spectral state transition is evident along the lightcurve decay of the source. With the drop in total count rate, the spectrum suddenly hardened while the cutoff energy of the power law component increased sharply. This could be explained in light of the 'propeller effect' (Illarionov, Sunyaev 1975). As the mass accretion rate drops during the decay process, the neutron star magnetosphere expands. When its radius $R_m$ exceeds the corotation radius $R_c$, the accreted material will not overcome the centrifugal barrier and get spun out. Therefore, the rate of mass accretion onto neutron star surface drops suddenly and significantly. When fewer soft photons are radiated into the hot plasma in the corona, their cooling of the electrons will be less effective. Therefore, we see a sudden increase in cutoff energy of the power law component, which has generally been regarded as closely related to the electron temperature in the Comptonization model.

More evidence for this explanation comes from analysis of the flux change of different spectral components. Across the transition, flux of the blackbody component dropped by almost a factor of 10 while the powerlaw flux and total flux changed only gradually. If we take bolometric luminosity as the total gravitational energy loss of the accreted material, the gap in total luminosity across the centrifugal barrier should be (Corbet, 1996; Campana & Stella, 2000)

$$\varDelta = 3 \, P_{2.5ms}^{2/3} \, M_{1.4}^{1/3} \, R_6^{-1}$$

where $P_{2.5\,ms}$ is neutron star spin period in units of 2.5 ms, $M_{1.4}$ is the mass of the neutron star in units of 1.4 solar mass, $R_6$ is the radius of the neutron star in units of 10 km. Taking the neutron star in 4U 1608-52 to be of 1.4 solar mass, 10 km in radius, and with a 1.61 ms spin period (see section 1 for more details), $\varDelta$ is only about 2, consistent with our data. The sudden decline in blackbody flux shows that a large fraction of the accreted material never reached the neutron star surface, but instead got propelled at the magnetospheric radius; the power-law component, assumed to be generated in the boundary region between the magnetosphere and the truncated accretion disk (Zhang et al, 1998), experienced only gradual decline because the amount of matter transferred to this region did not change significantly..

The constancy of blackbody temperature across the transition lends further support for this propeller scenario. When the accreted material falls in and eventually onto the neutron star surface, its gravitational energy loss is converted into radiation. Though the total amount of accretion dropped suddenly, the energy converted per unit mass does not change, thus the blackbody temperature.

Having identified the observed spectral state transition as the start of the propeller regime, inference about the neutron star surface magnetic field can be made in the following way. As the transition occurs when $R_m = R_c$, the critical overall luminosity at the point of the transition is related to the neutron star surface magnetic field as (Zhang et al, 1998):

$$L{\rm x},36 \approx 2.34\, B_9^2\, P_{-2}^{-7/3}\, M_{1.4}^{-2/3}\, R_6^5,$$

where $L{\rm x},36$ is the total X-ray luminosity in units of $10^{36}\, ergs\ s^{-1}$, $B_9$ the NS surface magnetic field in units of $10^9$ G, $P_{-2}$ the NS spin period in units of 10 ms. The total X-ray luminosity in observations 4 and 5 are $5.3 \times 10^{36}\ ergs\ s^{-1}$ and $3.3 \times 10^{36}\ ergs\ s^{-1}$ respectively, between which the transition luminosity lies. Assuming again the neutron star of 1.4 solar mass and 10 km in radius with a 1.61 ms spin period, the inferred magnetic field is $(1.4\text{-}1.8) \times 10^8$ Gauss, in agreement with the assumption that atoll sources have lower magnetic field ($\sim 10^8$ G) compared with Z sources ($> 10^9$ G).

3.2 Similarities and differences between spectral states in NS and BH systems

High energy spectra of BH binary systems usually fall into two categories (Zhang et al, 1997): a soft state with higher soft X-ray component luminosity, larger power law index (2.5-3.5) for the hard

X-ray component and no detectable break up to 300-600 keV; a hard state with lower soft X-ray luminosity, smaller power law index (1.5-2.0) and cutoff energy at 50-200 keV. In LMXB neutron star systems, similar spectral states show up and are called banana state and island state respectively (Campana et al, 2004). Type-I X-ray bursts have been observed in 4U 1608-52, so it is confirmed to harbor a neutron star instead of a black hole. For observations 1-4 in our analysis above, the flux is high and the spectrum is soft. As the total flux drops, the spectrum becomes harder in observations 5-19. This resemblance between BH and NS systems has encouraged efforts to propose similar models for state transition in both systems, which attribute the observed phenomena mainly to the accretion process (Esin et al, 1997; Menou et al, 1999).

There are still differences, though. In our observation, there seems to be an anti-correlation between total luminosity and cutoff energy of hard X-ray component. In the soft state, the cutoff is at about 3 keV while in the hard state, it jumps to above 30. When the flux is extremely low, as in observations 17-19, the upper limit of this parameter exceeded the upper limit of our detection range, which means that there is in fact no detectable spectral break. Therefore, we tried to fit the spectra data of observation 17-19 with model BB+PL and got satisfactory result (table 2). This is not consistent with the scenario in the BH system, where spectral break is evident in its hard state and disappears when the luminosity is high.

Table 2
Spectral Fitting Results of observations 17-19 with the BB+PL model

| Observation No. (1) | Power law | | Blackbody | | Iron line | | | $\chi^2/48$ (9) |
|---|---|---|---|---|---|---|---|---|
| | Index (2) | Norm (3) | kT (4) | Norm (5) | LineE (6) | Sigma (7) | Norm (8) | |
| 17 | $1.81^{+0.04}_{-0.04}$ | $0.122^{+0.010}_{-0.008}$ | $0.74^{+0.07}_{-0.09}$ | $32.63^{+11.07}_{-9.04}$ | $6.920^{+0.23}_{-0.26}$ | $<0.64$ | $4.77E-04^{+3.83E-04}_{-1.81E-04}$ | 0.948 |
| 18 | $1.54^{+0.05}_{-0.05}$ | $2.69E-02^{+0.44E-02}_{-0.50E-02}$ | $0.73^{+0.09}_{-0.11}$ | $21.28^{+6.97}_{-6.10}$ | $7.16^{+0.53}_{-0.59}$ | $0.85^{+0.71}_{-0.38}$ | $4.01E-04^{+3.72E-04}_{-2.01E-04}$ | 0.841 |
| 19 | $1.98^{+0.04}_{-0.03}$ | $0.159^{+0.005}_{-0.005}$ | $0.80^{+0.07}_{-0.11}$ | $15.37^{+4.75}_{-4.70}$ | $6.97^{+0.19}_{-0.18}$ | $<0.62$ | $5.31E-04^{+2.17E-04}_{-1.32E-04}$ | 0.778 |

Table 2. Fitting results of the model Bbodyrad+Powerlaw+Gaussian for the last three spectra. PCA data between 3.0-25.0 keV and HEXTE data between 15.0-150.0 keV are used. *NH* is fixed at $1.5 \text{ E } 22 \text{ cm}^{-2}$. Col. (1): Observation number. Col. (2): power law index. Col. (3): units of photons $\text{keV}^{-1} \text{cm}^{-2} \text{s}^{-1}$. Col. (4): units of keV. Col. (5): units of photons $\text{keV}^{-1} \text{cm}^{-2} \text{s}^{-1}$. Col. (6): units of keV. Col. (7): units of keV. Col. (8): units of photons $\text{keV}^{-1} \text{cm}^{-2} \text{s}^{-1}$. Col. (9): reduced chi-squared for 48 degrees of freedom.

Such differences suggest that the central accretor may play an important role in determining details of the spectral states. A solid surface of Neutron Stars may have a great impact on how energy is released. Large magnetic field on their surface might control the flow of accreted materials and dictate different stages of accretion, just like in this 'propeller' instance.

Acknowledgements:

X.C. would like to thank J. L. Qu, D. Lai, N.N. Tang, L. Shao for their advice and discussions which helped greatly with this work. SNZ acknowledges partial funding support by the Ministry of Education of China, Directional Research Project of the Chinese Academy of Sciences and by the National Natural Science Foundation of China under project no. 10521001.